# A Comparison of the Velocity Parameters of SiO v=1, J=1-0 and J=2-1 Maser Emission in Long Period Variables

## Abstract

We present an analysis of velocity parameters derived from multi-epoch observations of the SiO maser spectra of 47 long period variables (LPVs). The velocity parameters are important to inform and constrain theoretical models of SiO maser emission and to extract information on binary orbits. Mira and R Aquarii (R Aqr) are two known binaries included in the program. The 47 LPVs are among 121 sources of the Australia Telescope National Facility (ATNF) Mopra telescope's monitoring program. Observations were carried out several times a year between 2008 and 2012 and are continuing. The SiO spectra are from the v=1, J=1-0 (43.122 GHz; hereafter J10) and the v=1, J=2-1 (86.2434 GHz; hereafter J21) transitions. For 41 of the 47 LPVs we observed both transitions nearly simultaneously in 457 observations. We have determined and compared the velocity centroids (VCs) and velocity ranges of emission (VRs) suffixed as above (10 and 21) for the two transitions - VC10, VC21, VR10, and VR21. The VCs of the two transitions are, on average, within 0.13 km s$^{-1}$ of each other but are sometimes separated by a few km s$^{-1}$. The VC10s are, on average, slightly more positive than the VC21s. The values of the VCs in the two transitions have been compared to justify using both of these transitions to extract binary star orbital parameters. The arithmetic mean VR10 derived from 635 observations of 47 sources is 6.4 km s$^{-1}$ with a standard deviation of 3.4 km s$^{-1}$ while the mean VR21 derived from 485 observations of 41 sources is 4.2 km s$^{-1}$ with a standard deviation of 2.8 km s$^{-1}$. The number of occurrences of VR10 and VR21 versus velocity range have different distributions. The differences in the VRs indicate that the J21 and J10 emissions arise from dynamically different regions of the circumstellar environment.

## 1. Background

### 1.1 VCs and VRs

The VC can be thought of as the "center of mass" of the maser emission. The VC tracks the motion of the emission. It is assumed that, over time, the changes in the VC of the maser emission track the changes in the velocity of the actual center of mass of the stellar system. McIntosh (2006) showed that the VC of SiO maser emission varies with a standard deviation of 1.5 km s$^{-1}$ to 2 km s$^{-1}$ over time for LPVs not thought to be in binary sytems. Variations in excess of this standard deviation may indicate asymmetries, circumstellar or planetary orbits, the binary nature of the system, or other interesting physics.

Mathematically, the VC is the sum of the antenna temperature ($T_a$) in each velocity channel times the velocity with respect to the local standard of rest ($v_{lsr}$) of the velocity channel over the range of emission divided by the sum of the $T_a$ in each velocity channel over the range of emission.

$$VC = \frac{\Sigma (T_a * v_{lsr})}{\Sigma T_a}$$

The summations extend over the range of emission.

The VR is calculated to be the region where the $T_a$ exceeds three times the standard deviation of the antenna temperature of the background noise. The standard deviation is determined from velocity channels far away from the emission range of the source, which, given the IF bandwidth of 137 MHz and resolution of 4096 channels is easily obtainable by masking the central emission.

Analysis and comparison of the VRs has received little attention. VRs and their phase dependence are commonly displayed in modeled spectra but values are not explicitly presented. VR similarities or differences among transitions provide information on the dynamics of the region in which the masers originate.

## 1.2 Binary Stars

The VCs of SiO masers and other molecular transitions have been used in attempts to determine the orbital parameters of R Aqr. R Aqr is an LPV, an SiO maser source, and a symbiotic star. Its period has been predicted from an analysis of the VCs of various molecules and transitions to be 35 years (McIntosh and Rustan 2007 hereafter MR07) or 44 years (Gromadski and Mikolajewska 2009 hereafter GM09).

MR07 and GM09 assumed that different transitions of the same or different molecules tracked the same stellar velocity. The Mopra data base has been analyzed to check this assumption with the J21 and J10 transitions. Simultaneous observations with the other transitions are not available to verify the assumption more generally.

## 1.3 Observations of Maser Locations and VRs

Phillips et al. (2003) simultaneously viewed the J21 and J10 transitions around R Cassiopeia. The J10 emission had a higher flux density and more observed features, but the VRs were 4 km s$^{-1}$ to 5 km s$^{-1}$ for both transitions. The results indicated that the two transitions "form at comparable radii from R Cas." Soria-Ruiz et al. (2004) observed X Cyg and concluded "the size of the emitting regions is comparable," and the autocorrelation spectra showed a VR10 of ~9 km s$^{-1}$ and VR21 ~12 km s$^{-1}$. Soria-Ruiz et al. (2004) found that for IRC +10011 the J21 emission was located "50% further away" from the assumed centroid of emission compared to the J10 emission. The autocorrelation spectra for IRC+10011 showed a VR10 of ~10 km s$^{-1}$ and VR21 ~9 km s$^{-1}$. A map of the emission from various transitions for IRC +10011 is found in Colomer et al. (2009). The J21 "shows a significantly different spatial distribution," and is further from the star than the J10 emission. Colomer et al. (2009) does not include any spectra. In TX Cam Soria-Ruiz et al. (2006) found that the angular size of the J21 region was larger than the J10 region. The velocity ranges of emission in the two transitions appear similar in the published spectra. For R Leo (Soria-Ruiz et al. 2007) the J21 emission is produced in a "clearly farther shell" than the J10 and the velocity range of the J10 exceeds that of the J21 emission. R Leo, the only source observed by VLBI in both J10 and J21 and included in the Mopra program, the autocorrelation spectra indicate a VR10 of ~16 km s$^{-1}$ and VR21 of ~12 km s$^{-1}$. These observations indicate that the distance of the J21 emission is comparable to or greater than the distance of the J10 emission from the central star, but do not indicate a clear relationship between VR10 and VR21.

## 1.4 Theoretical Models of Maser Velocity Parameters and Locations

Gray et al. (2009; hereafter G09) have investigated the dynamics of the circumstellar region in which the SiO masers originate and have provided the most thoroughly developed theory for the maser spectra in LPVs. They model and depict a shock traveling out from the star generating different velocities, red shifted or more positive and blue shifted or more negative, at different distances from the star. As the shock travels out the velocities change as a function of distance from the star and phase. G09 predicted a VR10 of ~10 km s$^{-1}$. From the graphs in G09 the VR21s are generally smaller than VR10s, but the difference is phase dependent and difficult to quantify. According to the information presented in G09 a redward shift of the spectral peak with increasing phase for J10 is predicted, but no phase averaged VC shift between the two transitions is obvious. G09 indicate that the J21 emission should originate at similar or greater distances from the star than the J10 emission depending on the phase.

Yun & Park (2012; hereafter YP12) have developed an SiO maser emission model for LPVs that takes into account the complex velocity field of the circumstellar region using a coupled escape probability. The graphs in this work indicate VR21s exceeding the VR10s at all epochs presented. No clear difference between VC21 and VC10 are indicated at any epoch. The J21 emission is generated at similar or slightly greater distances from the star than the J10 emission in this model.

Since different distances from the star are affected differently by the proposed shock travelling out from the star, it is reasonable to expect that masers forming at different distances from the star will exhibit different VRs and slightly different VCs. The observations of the velocity parameters of the emission provide information on the locations as well as the motion of the masing material.

## 2. Observations

The 22-m Mopra radio telescope is located in New South Wales, Australia. At Mopra each maser source is first pointed on to ascertain adequate positioning inside the beam. The observations are then executed as 16 cycles on and 16 cycles off observation, thus lasting 64 seconds per on/off pair. For J21 observations the rms noise was about 1.8 Jy. For J10 observations the rms noise was about 0.6 Jy. The velocity resolution was 0.23 km s$^{-1}$ (J10) and 0.12 km s$^{-1}$ (J21). VR10 averages 6.4 km s$^{-1}$ or about 28 velocity channels. VR21 averages 4.2 km s$^{-1}$ or about 34 velocity channels. The velocity resolution is therefore about 3% to 3.5% of the average velocity range and limits the accuracy of VCs and VRs to approximately 0.2 km s$^{-1}$.

The Mopra monitoring program observed 121 sources between 2008 and early 2012 and is continuing with a reduced rate of observations. Long period variables (LPVs), semi-regular variables (SRVs), irregular variables, OH-IR stars, and the Orion SiO maser source were observed approximately monthly in J10 and J21. Two of the LPVs, Mira and R Aqr, are known to be binary stars. It has been suggested that L2 Puppis, an SRV included in the Mopra program, is a binary star (Goldin & Makarov 2007)

LPVs were chosen for this analysis because they present a relatively uniform collection of sources and a group of stars for which theoretical models of the SiO maser emission have been developed. The 47 LPVs observed, epoch of the first simultaneous or J10 observation, the number of observations in each transition, and the number of observations of both transitions within 24 hours are given in Table 1. These sources are listed in the AAVSO Bulletin (2011) as LPVs with well determined maxima and periods.

## 3. Velocity Parameters and Results

Table 2 presents the VC10, VC21, VR10, and VR21 for each source determined at the first epoch of observation. Included in Table 2 is the number of times the magnitude of the difference between VC21 and VC10 is greater than 2 km s$^{-1}$. The number of occurrences of these large VC differences is indicated by the signs of the difference. Sources with occurrences of VR21 > VR10 are indicated by bullets.

## 3.1 VC Comparison

We determined the difference in velocity for 457 pairs of VC10s and VC21s from 41 LPVs. Six LPVs were not detected in J21. The VC pairs are from the same source and were obtained within 24 hours of each other. The arithmetic mean of the differences between the VC21 and the VC10 is -0.1 km s$^{-1}$ with a standard deviation (square root of the sample variance) of 1.1 km s$^{-1}$.

Figure 1 displays the histogrammed results of VC21 – VC10. The distribution is not particularly well described by the Gaussian fit to the histogrammed data. The poor fit is most obviously indicated by the number of occurrences beyond -2 km s$^{-1}$ compared to the Gaussian fit. The Gaussian fit maximum occurs at +0.1 km s$^{-1}$ and the fit Gaussian sigma is 0.8 km s$^{-1}$.

The poor fit may exist because the velocity parameters at one observations of a source are related to the parameters observed a few weeks to a few months later. Gaussian distributions apply to randomly occurring events. McIntosh and Bonde (2010) determined SiO maser features for Mira to have an average lifetime of 171 days. This lifetime is much longer than the average time between measurements of the Mopra program.

The VC differences displayed in Figure I are not symmetric about the average value. The differences indicate that more VC21s are within about 1 km s$^{-1}$ of the average while many VC10s are red shifted several km s$^{-1}$ from the average difference. There may be an unknown physical reason for the J10 emission to be more redshifted than the J21 emission. For individual sources the VC21 and VC10 may differ by several km s$^{-1}$ at specific epochs.

Since VC10s generally have a standard deviation of 1.5 km s$^{-1}$ to 2.0 km s$^{-1}$ over time (McIntosh 2006), the observations that the VCs of the two transitions are usually within approximately 1 km s$^{-1}$ of each other justifies combining VC10s and VC21s to track changes in the VCs of sources. GM09 made this assumption in fitting the velocity curve of R Aqr to extract the binary parameters.

Knowing the general variations of VCs, individual source spectra can be examined for the existence of periodic variations. It has been suggested (Struck et al. 2002; Cohanim 2002) that planets may orbit LPVs in the circumstellar environment and affect the SiO maser emission. Such planets, if they exist, should be detectable through long term studies of maser VCs, VRs, or the motions of individual maser features. Presently no observational evidence exists to support the presence of these planets.

In future work we will analyze the VCs of the Mira and R Aqr emission over time, including these Mopra data, to determine their binary parameters. We will also search the VCs of other program sources for periodicities indicating binary partners, planets, or other interesting physics.

Four sources, R Hor, RS Vir, RR Aql, and S Gru, the VC21 – VC10 difference has a negative value and a magnitude greater than 2 km s$^{-1}$ on at least three epochs as indicated in Table 2. Only one star, R Cnc, shows VC21 – VC10 > +2 km s$^{-1}$ at more than one epoch. The large VC differences often indicate that features are present in the one transition that are not present in the other transition. The reason for the predominance of red shifted VC10s with respect to the VC21s is not known. Future observations may determine if these large differences are chronic conditions or short term statistical variations. If the large VC differences continue, VLBI observations of these sources might indicate an unusual geometry or location of the maser features.

As stated above the VC10s are generally more positive (red shifted) than the VC21s. It is difficult to compare this result, even qualitatively, with the G09 and YP12 models. The G09 modeled spectra indicate a blue shifted J10 emission compared to the J21 emission for some combinations of phase and dust regime. Other combinations produce a relative red shift of J10 compared to J21. The graphs in YP12 indicate that the VCs are approximately equal for all epochs presented. The Mopra spectra will have to be analyzed as a function of stellar phase in order to make a more detailed comparison with the predictions.

### 3.2 VR Comparison

The arithmetic mean of the VR10s is 6.4 km s$^{-1}$ with a standard deviation of 3.4 km s$^{-1}$. Figure 2 shows the histogrammed VR10s. The number of occurrences versus VRs in 1 km s$^{-1}$ windows has been fit by a Poisson distribution (Bevington & Robinson 1992) . The Poisson distribution is defined by μ, the mean of the Poisson distribution, and σ, the standard deviation of the Poisson distribution. This distribution is often used to model histogrammed data with a relatively small number of data points. Assuming a Poisson distribution, the μ of the VR distribution is 5.6 km s$^{-1}$ with a σ of 2.4 km s$^{-1}$. The peak number of occurrences is in the 2 km s$^{-1}$ to 3 km s$^{-1}$ range although the distribution is fairly flat from 2 km s$^{-1}$ through 6 km s$^{-1}$. Again the Poisson distribution does not fit the data very well. This poor fit again may be affected by the non-random relationship of the emission of a source from one observation to another. Both the VR10 arithmetic mean and the Poisson fit μ are considerably smaller than the ~10 km s$^{-1}$ stated by G09 for the VR10.

The arithmetic mean of the VR21s is 4.2 km s$^{-1}$ with a standard deviation of 2.8 km s$^{-1}$. Figure 3 shows the histogrammed VR21s. The fit Poisson distribution indicates a μ of 2.8 km s$^{-1}$ with a σ 1.7 km s$^{-1}$. The peak number of occurrences is in the 1 to 2 km s$^{-1}$ range.

The results shown in Figs 2 and 3 indicate that the J10 emission occurs over a larger VR than the J21 emission. This result is qualitatively consistent with the predictions of G09. The modeled spectra presented in G09 indicate a VR10 that exceeds the VR21 by approximately 1 km s$^{-1}$ at all phases and for all dust regimes.

The YP12 modeled spectra show the VR21 as larger than the VR10 for all epochs and is not consistent with the observations.

Figure 4 shows the range of VR21 versus the VR10 for simultaneous observations. For only 17 of the 457 simultaneous observations does the VR21 exceed the VR10. The sources and the number of times VR21 exceeded VR10 are indicated in Table 2. So on average, and for almost all individual, simultaneous, pairs of observations the VR10 exceeds the VR21.

The difference in VR10 and VR21 may be associated with a difference in location or the range of possible locations of the masers in the two transitions. The J21 emission appears to occur in a region of the circumstellar envelop that generates features moving at a smaller velocity with respect to the central star. The limited VLBI data available indicate that the J21 emission arises further from the star than the J10 emission at least in some cases. The J21 has not been observed closer to the star.

## 4. Future Work

In further analysis of the database we will focus on the VR and VC as a function of phase for these LPVs. The phase dependent analysis will allow a much more complete comparison with the predictions of G09 and YP12. The variations of the VCs of individual sources will also be analyzed to look for evidence of binarity or periodic changes in the velocity structure of the source.

The phase relationship between the VRs of different transitions has not been previously investigated. A previously existing data set of the $v = 0, 1,$ and $2; J = 1-0$ transitions (McIntosh 2006), along with the Mopra data set, can now be examined for correlations among the transitions, VRs, and stellar phase. Such correlations may provide evidence for the relative locations of the masers in different transitions and the motion of the proposed shocks through the maser region.

We will search the acquired spectra for periodic variations in VCs, VRs, and features. Such variations could indicate binarity, planets, or other interesting physical changes.

## 5. Conclusions

The Mopra database provides the first large data set of LPV SiO maser spectra (essentially simultaneous observations in J21 and J10 over the last 4 years) to allow the comparison of the VCs and VRs for the J10 and J21 transitions. The velocity comparisons extracted from these observations will inform and constrain the development of future models of the circumstellar environment and maser dynamics.

The analysis of simultaneously acquired J10 and J21 VCs justify the assumption that these values can both be used in the extraction of binary parameters. The average difference in the VCs of these two transitions is very small compared to the time variation of the VCs.

The J10 emission shows a larger VR than the J21 emission. This result indicates that the two transitions originate in different parts of the circumstellar envelope and experience the proposed shock propagation differently. The larger VR10 is qualitatively consistent with the predictions of G09. However a detailed comparison of the predicted and observed J10 and J21 VCs and VRs is not possible at this point.

## 6. References

AAVSO Bulletin 74: Predicted Dates of Maxima and Minima of Long Period Variables 2011, http://www.aavso.org/aavso-bulletin

Bevington, P. & Robinson, D. 1992, Data Reduction and Error Analysis for the Physical Sciences, 2$^{nd}$ edition, (McGraw-Hill, New York)

Cohanim, B. 2002, *JAAVSO*, **30**, 145

Colomer, F. 2009, Proceedings of "The 9th European VLBI Network", eds Mantovani et al., (Bologna, Italy; Proceedings of Science)

Goldin, A. & Makarov, V. 2007, *ApJSS*, **173**, 137

Gray, M. D., Wittkowski, M., Scholz, M., Humphreys, E. M. L., Ohnaka, K., & Boboltz, D. 2009, *MNRAS*, **394**, 51 (G09)

Gromadski, M. & Mikolajewska, J. 2009, *A&A*, **495**, 931 (GM09)

McIntosh, G. and Bonde, J. 2010, PASP, **122**, 396

McIntosh, G. and Rustan, G. 2007, *AJ*, **134**, 2113 (MR07)

McIntosh, G. 2006, *AJ*, **132**, 1046

Phillips, R.B., Straughn, A.H., Doeleman, S.S., & Lonsdale, C. J. 2003, *ApJ*, **588**, L105

Soria-Ruiz, R., Alcolea, J., Colomer, F., Bujarrabal, V., & Desmurs, J.-F. 2007, *A&A*, **468,** L1

Soria-Ruiz, R., Colomer, F., Alcolea, J., Bujarrabal, V., Desmurs, J. F., & Marvel, K. B. 2006, Proceedings of the 8th European VLBI Network Symposium,Torun, Poland. Eds : Baan et al.

Soria-Ruiz, R., Alcolea, J., Colomer, F., Bujarrabal, V., Desmurs, J.-F.. Marvel, K. B., & Diamond, P. J. 2004, *A&A*, **426**, 131

Struck, C., Cohanim, B., & Willson, L. 2002, *ApJ*, **572**, L83

Yun, Y., & Park, Y-S. 2012, *A&A*, **545**, 136

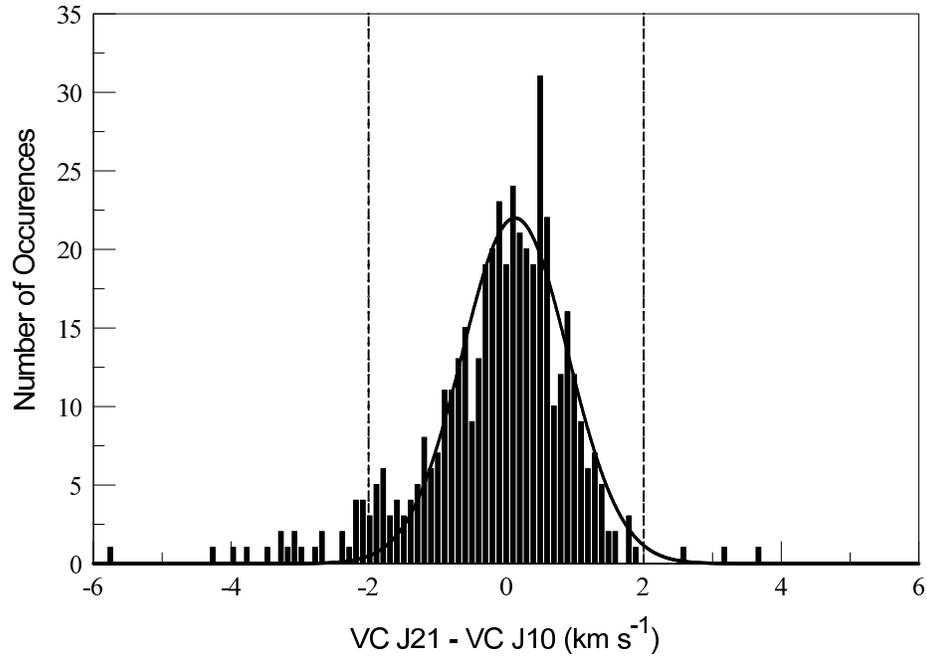

Figure 1. Histogram of VC21 – VC10 for observations within one day of each other from the Mopra data base for 457 J21 and J10 spectra pairs. The solid line is the Gaussian fit to the histogrammed data. The dashed vertical lines are drawn at +2 km s$^{-1}$ and -2 km s$^{-1}$.

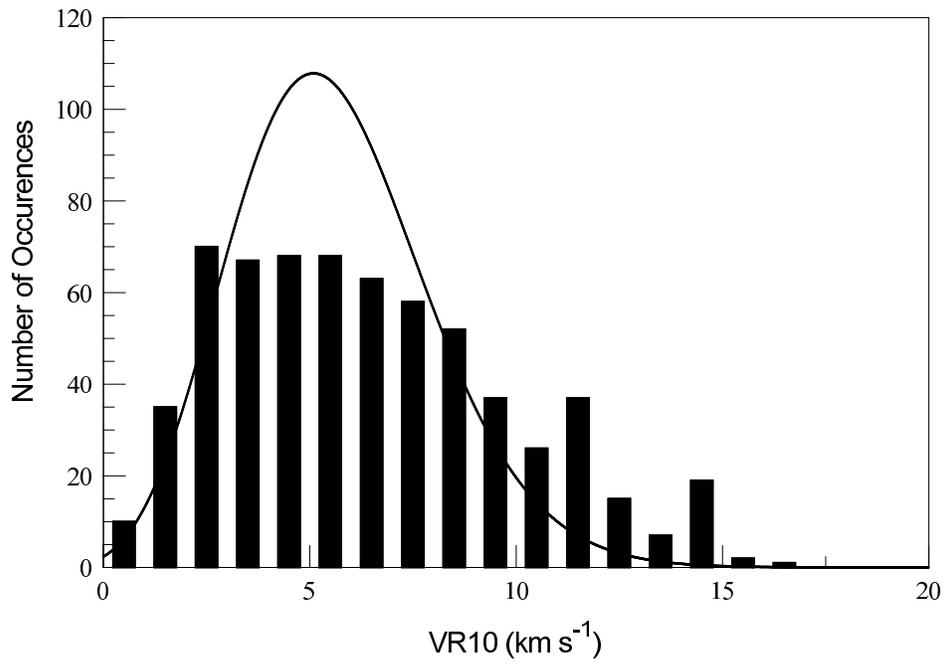

Figure 2. Number of occurrences of VR10 versus the range of emission from the Mopra sources. There are 635 J10 observations included. The solid line is the Poisson fit to the histogrammed data.

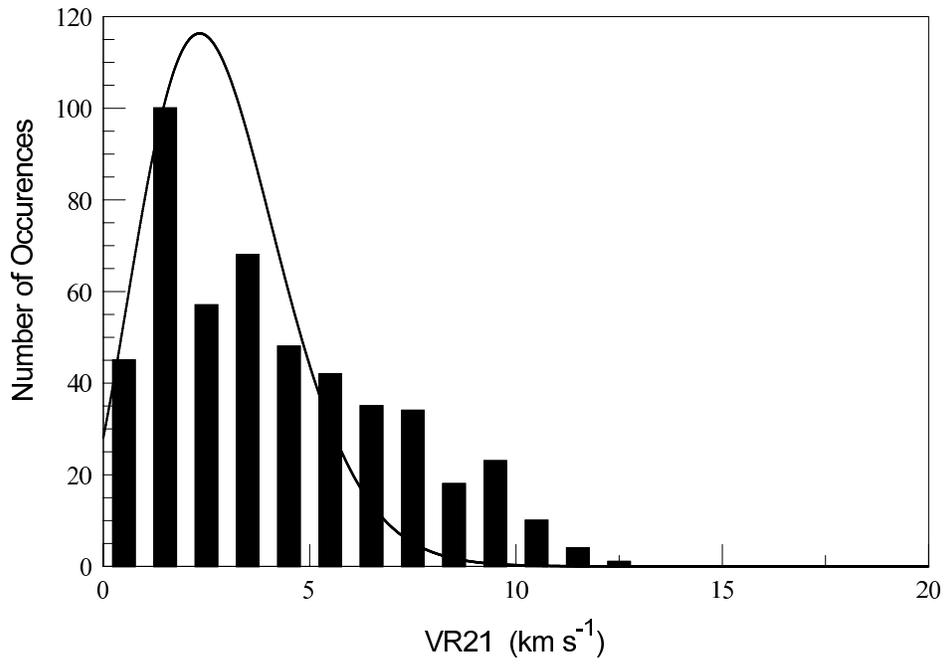

Figure 3. same as Figure 2 for 485 J21 observations

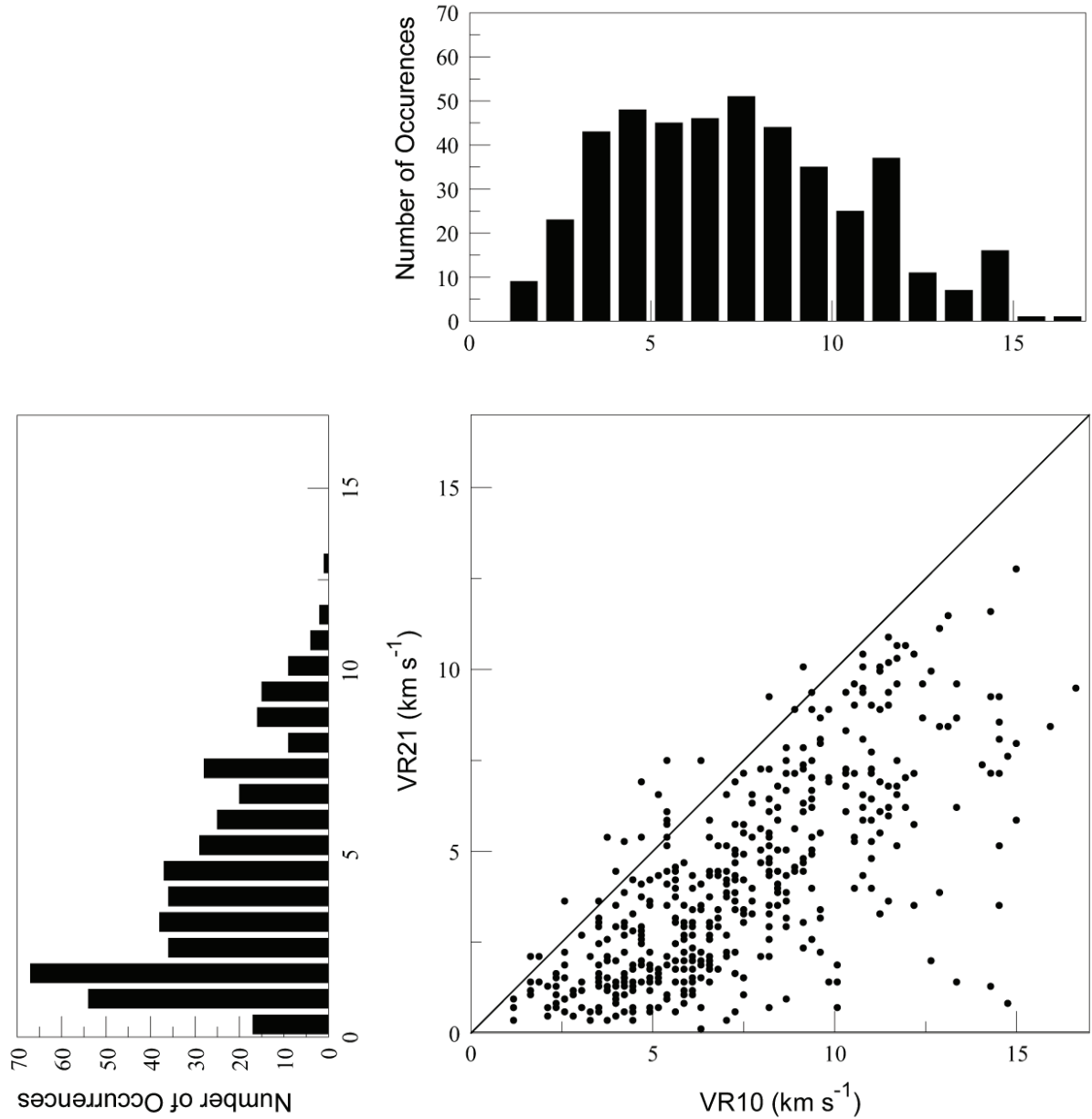

Figure 4. VR21 versus VR10 for simultaneous observations. The solid line represents equal velocity ranges in the two transitions. Included in this plot are 457 simultaneous observations. The 17 observations above the line are indicated in Table 2. The top histogram indicates the VR10 values. The left histogram indicates the VR21 values.

**Table 1**
**Sources and number of observations**

| Source | Epoch of First J10 or Simultaneous Observation | J10 Observations | J21 Observations | Simultaneous Observations |
|---|---|---|---|---|
| Z Peg | 2010 Oct 21 | 5 | 0 | 0 |
| S Scl | 2009 Apr 15 | 19 | 15 | 14 |
| R Psc | 2011 Jan 12 | 3 | 0 | 0 |
| **Mira** | 2009 Oct 5 | 20 | 19 | 16 |
| R Cet | 2010 Jan 19 | 10 | 0 | 0 |
| R Hor | 2009 Oct 5 | 18 | 15 | 14 |
| W Eri | 2010 Mar 16 | 12 | 11 | 10 |
| R Tau | 2011 Aug 13 | 14 | 3 | 3 |
| R Cae | 2009 Apr 21 | 16 | 13 | 13 |
| T Lep | 2009 Oct 5 | 14 | 14 | 14 |
| S Pic | 2009 Apr 22 | 18 | 18 | 18 |
| R Oct | 2009 Oct 5 | 16 | 15 | 15 |
| S Ori | 2009 Apr 21 | 21 | 18 | 18 |
| S Col | 2010 Jan 19 | 16 | 15 | 15 |
| U Ori | 2009 Oct 5 | 19 | 22 | 19 |
| R Cnc | 2009 Apr 28 | 17 | 17 | 16 |
| W Cnc | 2009 Apr 28 | 9 | 1 | 1 |
| R Car | 2009 Oct 5 | 12 | 11 | 11 |
| X Hya | 2009 Apr 28 | 10 | 9 | 8 |
| R LMi | 2009 Oct 5 | 16 | 14 | 13 |
| R Leo | 2009 Apr 28 | 18 | 20 | 18 |
| W Leo | 2011 Dec 1 | 3 | 0 | 0 |
| X Cen | 2010 Jan 19 | 15 | 6 | 6 |
| T Vir | 2011 Oct 14 | 2 | 0 | 0 |
| R Hya | 2009 Apr 28 | 20 | 20 | 20 |
| S Vir | 2009 Apr 28 | 19 | 11 | 10 |
| RU Hya | 2011 Aug 15 | 7 | 2 | 2 |
| RS Vir | 2010 Apr 12 | 14 | 8 | 8 |
| S Ser | 2010 May 27 | 15 | 2 | 2 |
| RS Lib | 2009 Oct 6 | 15 | 10 | 10 |
| R Ser | 2009 Apr 28 | 14 | 6 | 5 |
| RU Her | 2010 Apr 12 | 15 | 7 | 6 |
| U Her | 2009 Oct 5 | 17 | 18 | 16 |
| T Oph | 2010 Apr 12 | 12 | 9 | 8 |
| X Oph | 2009 Oct 5 | 18 | 17 | 17 |
| R Aql | 2009 Oct 5 | 16 | 15 | 13 |
| RT Aql | 2009 Oct 5 | 14 | 7 | 7 |
| S Pav | 2009 Oct 5 | 13 | 11 | 10 |
| RR Sgr | 2010 Mar 15 | 10 | 8 | 8 |
| RR Aql | 2009 Oct 5 | 15 | 11 | 11 |
| RU Cap | 2010 Apr 12 | 2 | 3 | 2 |
| W Aqr | 2009 Oct 5 | 13 | 10 | 10 |
| V Peg | 2011 Oct 12 | 1 | 0 | 0 |
| S Gru | 2009 Oct 5 | 8 | 6 | 6 |
| R Peg | 2009 Apr 20 | 18 | 13 | 12 |
| W Peg | 2009 Apr 20 | 17 | 14 | 14 |
| **R Aqr** | 2009 Apr 20 | 19 | 21 | 18 |

Notes. Known binary stars are identified in **bold**. Details of the individual observations and spectra can be found at http://www.narrabri.atnf.csiro.au/cgi-bin/obstools/siomaserdb.cgi

**Table 2**
**Sources, VCs, and VRs**

| Source | VC10 (km s$^{-1}$) | VR10 (km s$^{-1}$) | VC21 (km s$^{-1}$) | VR21 (km s$^{-1}$) | ΔVC> \|2 km s$^{-1}$\| | VR21>VR10 |
|---|---|---|---|---|---|---|
| Z Peg  | -27.1 | 2.1  |       |     |         |     |
| S Scl  | 3.3   | 2.3  | 6.9   | 0.7 | +       |     |
| R Psc  | -57.8 | 0.7  |       |     |         |     |
| *Mira* | 49.1  | 4.2  | 47.9  | 2.9 | -       | ●   |
| R Cet  | 33.7  | 2.1  |       |     |         |     |
| *R Hor*| 40.4  | 8.1  | 38.6  | 5.4 | - - - - - - |   |
| W Eri  | -0.9  | 2.3  | -0.6  | 1.5 |         |     |
| R Tau  | 16.0  | 4.9  | 15.8  | 1.9 |         |     |
| R Cae  | 0.4   | 9.4  | 0.7   | 7.5 | -       |     |
| T Lep  | -28.7 | 4.9  | -29.2 | 1.8 |         |     |
| S Pic  | -2.2  | 6.6  | -2.6  | 4.4 |         | ●   |
| R Oct  | 19.3  | 6.1  | 19.4  | 3.4 | -       |     |
| S Ori  | 10.9  | 9.1  | 11.9  | 3.0 |         | ●●● |
| S Col  | 63.0  | 4.7  | 62.9  | 2.9 |         |     |
| U Ori  | -37.6 | 6.6  | -38.1 | 5.8 |         |     |
| *R Cnc*| 14.0  | 8.2  | 12.8  | 3.3 | + +     |     |
| W Cnc  | 37.5  | 3.7  | 33.3  | 0.4 | -       |     |
| R Car  | 10.6  | 12.9 | 10.0  | 3.9 |         | ●●  |
| X Hya  | 26.7  | 4.4  | 27.6  | 4.2 |         | ●   |
| R LMi  | 0.0   | 12.2 | 0.4   | 3.5 |         |     |
| R Leo  | -0.6  | 5.4  | -2.7  | 7.5 | - -     | ●●  |
| W Leo  | 43.1  | 4.4  |       |     |         |     |
| X Cen  | 28.6  | 1.2  | 29.2  | 0.7 |         |     |
| T Vir  | 5.2   | 2.3  |       |     |         |     |
| R Hya  | -9.4  | 8.7  | -11.1 | 5.0 |         | ●   |
| S Vir  | 9.8   | 4.7  | 10.4  | 2.6 |         |     |
| RU Hya | -2.9  | 6.1  | -2.9  | 2.3 |         |     |
| *RS Vir*| -10.7| 4.0  | -13.4 | 0.5 | - - - - |     |
| S Ser  | 21.1  | 4.2  | 21.3  | 0.6 |         |     |
| RS Lib | 9.6   | 8.0  | 8.0   | 2.1 | -       | ●   |
| R Ser | 34.5 | 4.7 | 33.7 | 1.4 | | |
| RU Her | -12.8 | 13.3 | -14.1 | 1.4 | | |
| U Her | -15.3 | 8.2 | -14.8 | 4.3 | | |
| T Oph | -31.9 | 5.6 | -32.2 | 1.8 | | ● |
| X Oph | -58.0 | 6.3 | -57.1 | 4.1 | | |
| R Aql | 45.3 | 6.6 | 45.4 | 2.7 | | ● |
| RT Aql | -30.7 | 5.2 | -30.4 | 1.5 | - | |
| S Pav | -22.4 | 4.9 | -23.3 | 3.5 | | ● |
| RR Sgr | 95.5 | 3.7 | 94.0 | 1.2 | - | |
| *RR Aql* | 28.6 | 8.7 | 28.7 | 0.9 | - - - - | |
| RU Cap | 6.6 | 2.3 | 6.9 | 1.1 | | |
| W Aqr | 1.1 | 2.6 | 1.5 | 1.9 | | |
| V Peg | -18.2 | 1.2 | | | | |
| *S Gru* | -27.1 | 1.2 | -26.8 | 0.4 | - - - | |
| R Peg | 21.8 | 6.6 | 22.0 | 3.0 | | |
| W Peg | -12.5 | 7.3 | -14.4 | 1.6 | | ● |
| R Aqr | -23.9 | 11.7 | -23.4 | 9.6 | | ● |

Notes. Sources with multiple observations of ΔVC> $|2$ km s$^{-1}|$ are identified in *italics*.
Sources observed to have VR21>VR10 are underlined.